# Pipeline for Automating Compliance-based Elimination and Extension (PACE²): A Systematic Framework for High-throughput Biomolecular Material Simulation Workflows

*Srinivas C. Mushnoori[1], Ethan Zang[1], Akash Banerjee[1], Mason Hooten[2], Andre Merzky[3], Matteo Turilli[3], Shantenu Jha[3] and Meenakshi Dutt[1]\*\**

1 - Chemical and Biochemical Engineering, Rutgers, The State University of New Jersey, Piscataway, New Jersey 08854

2 - Biomedical Engineering, Rutgers, The State University of New Jersey, Piscataway, New Jersey 08854

3 - Electrical and Computer Engineering, Rutgers, The State University of New Jersey, Piscataway, New Jersey 08854

\*\* corresponding author: meenakshi.dutt@rutgers.edu

## ABSTRACT

The formation of biomolecular materials via dynamical interfacial processes such as self-assembly and fusion, for diverse compositions and external conditions, can be efficiently probed using ensemble Molecular Dynamics. However, this approach requires a large number of simulations when investigating a large composition phase space. In addition, there is difficulty in predicting whether each simulation is yielding biomolecular materials with the desired properties or outcomes and how long each simulation will run for. These difficulties can be overcome by rules-based management systems which include intermittent inspection, variable sampling, premature termination and extension of the individual Molecular Dynamics simulations. The automation of such a management system can significantly reduce the overhead of managing large ensembles of Molecular Dynamics simulations. To this end, we propose a computational framework – the Pipeline for Automating Compliance-based Elimination and Extension (PACE²), for high-throughput ensemble biomolecular materials simulations. The PACE² framework encompasses

Simulation-Analysis Pipelines, where each Pipeline includes temporally separated simulation and analysis tasks. When a Molecular Dynamics simulation completes, an analysis task is triggered which evaluates the Molecular Dynamics trajectory for compliance. Compliant Molecular Dynamics simulations are extended to the next Molecular Dynamics phase with a suitable sample rate to allow additional, detailed analysis. Non-compliant Molecular Dynamics simulations are eliminated, and their computational resources are either reallocated or released. The framework is designed to run on local desktop computers and high performance computing resources. We present preliminary scientific results enabled by the use of PACE$^2$ framework, which demonstrate its potential as well as serve to validate PACE$^2$. In the future, the framework will be extended to address generalized workflows and investigate composition-structure-property relations for other classes of materials.

## INTRODUCTION

The formation of biomolecular materials via dynamical processes such as self-assembly or fusion of interfaces is key to many subdisciplines in life sciences and engineering. [1] The characteristics of the materials depend on the chemistry and concentration of the constituent biomolecules. Some synthetic chemistry approaches may identify relationships between the composition and characteristics of biomolecular materials, but they can be expensive, time consuming and have limited spatiotemporal resolution to elucidate dynamical processes underlying the formation of these materials. A large array of Molecular Dynamics simulations is better suited to investigating the dynamical processes underlying the formation of biomolecular materials under a variety of compositions and external conditions. This approach is an example of "ensemble Molecular Dynamics". [2]

The study of the formation of biomolecular materials using ensemble Molecular Dynamics suffer from two challenges: 1) many independent ensembles are necessary to adequately probe the composition phase space [3] and 2) it is difficult to predict how much simulation time will be needed for the desired materials to form. [4] Actual simulation runtimes are of interest because they provide a measure of the computational cost of those simulations. The cost of computation is familiar to many simulation scientists since state of the art Molecular Dynamics simulations are frequently run on high performance clusters, a significant financial investment. Hence, a computational framework which addresses these challenges and manages the efficient use of computational resources is needed.

An attentive computational scientist will strive to manage the length of each candidate simulation in the ensemble by adopting several management strategies. These management strategies include: A) intermittent inspection of simulations for signs of the formation of the desired biomolecular material; B) sparse sampling of simulations before they show evidence of formation, and frequent sampling thereafter; C) premature termination of simulations that evolve contrary to the reference analytical calculations; or D) extension of simulations which terminate without clear results. [4]

With these strategies, the computational scientist may decrease the duration of unproductive simulations, and therefore decrease the average cost per candidate simulation. The implementation of such strategies involves significant organization on the part of the scientist. Furthermore, strategies A through D can be implemented as a rules-based management routine. In particular if candidate simulations can be automatically assessed for conformity using a user-defined rule, then the elimination or extension of candidates can be adaptively managed by software. Hence, the long-term management of an ensemble of Molecular Dynamics simulations can be replaced by a decision loop with programmed analysis and conditional extension or termination of individual Molecular Dynamics simulations.

To that end, a computational framework for ensembles of biomolecular simulation pipelines is proposed. Specifically, the computational framework encompasses a Simulation-Analysis Pipeline (see Figure 1). The notion of a Simulation-Analysis Pipeline is straightforward: A Pipeline carries sequential simulation and analysis tasks, and a control flow decision is made based on the signal received from the analysis task. In other words, the first Molecular Dynamics simulation runs to completion for a fixed number of timesteps and triggers an analysis task. The analysis task operates on the resultant Molecular

Dynamics trajectory to evaluate a user-defined compliance metric. Compliant Molecular Dynamics simulations are extended to the next Molecular Dynamics (MD) phase with a high sample rate to allow further, detailed analysis. Non-compliant simulations are eliminated, and their computational resources are either allocated elsewhere or released. Resource allocation consideration is especially important due to the limited availability of high-performance computing resources. The Pipeline can be replicated for concurrent processing of multiple candidates simultaneously.

In this manuscript, the implementation of the Pipeline for Automating Compliance-based Elimination and Extension (PACE$^2$) framework as a general-purpose design for the management of Molecular Dynamics ensembles is discussed. Also, two application case studies in which PACE$^2$ is leveraged to automate the generation of a phase diagram for biomolecular materials encompassing chemical species of interest are discussed. In the future, the PACE$^2$ framework will be extended to handle more generalized workflows. Furthermore, the framework has the potential to be integrated with other computational methods and analysis tools to enable the identification of composition-structure-property relations in other classes of materials such as soft, polymeric materials, metals, ceramics and glasses.

**METHODS**

*The PACE$^2$ Framework*

The process of solving large ensemble problems via a high-throughput simulation framework can be set up in the form of shell scripts that explicitly assigns resources to all the simulations in question and resolves the necessary dependencies. While possible, this approach can prove to become extremely cumbersome, especially over large candidate pools (see Table A.1).[3,4] A large candidate pool requires the explicit specification of command lines referring to each system. This introduces the notion of a *task graph*, a complex set of tasks with interconnected dependencies. A simulation scientist may need to run Molecular Dynamics simulations (i.e., tasks) which may have multiple instrumental variables, start and stop at arbitrary points, dump simulation states to disk, or use these dumped states to set up and run new simulations (i.e., resolve dependencies). This points to the need for the workflows to be *adaptive:* i.e., have the ability to modify themselves during runtime in response to specific events that occur during simulations. Examples of events include failed simulations that require re-running, ensemble members that no longer need to be sampled and must be terminated, or ensembles that can safely terminate because they have reached a dynamically determined "desirable end-state." This allows for optimal use of precious compute time in addition to removing the need for human intervention and reducing the scope for human error. The expression of the adaptivity required in a workflow such as this is extremely limited when done via shell scripting, underscoring the requirement of a framework that provides advanced resource management capability and exposes them via an application programming interface (API), which allows for scalable and flexible expression of adaptive workflows. The description of the adaptability as an attribute of task graphs is discussed in a prior study.[5] Similarly, details of task graphs, workflows and adaptivity are provided in prior publications. [6,7]

The advanced resource management capabilities and an API to leverage it has been addressed by RADICAL Cybertools (RCT). RCT is a suite of middleware tools designed to enable the execution of multi-task applications in a scalable, flexible and extensible manner on a wide range of high performance computing (HPC) platforms. The RCT suite consists of three main modules: RADICAL-SAGA (Simple API for Grid Applications), RADICAL-Pilot, and RADICAL-EnTK (Ensemble Tool Kit). Put together, RADICAL-Cybertools allows the application to execute large ensembles of simulation-analysis tasks by enabling execution on HPC resources,[8] managing resource allocation and runtime details [9] and resolving the appropriate dependencies and containerizing tasks. [7]

These modules allow for an application to perform scalable, synchronous or asynchronous execution of large ensembles of tasks in any arbitrarily complex dependency order. The proposed framework is an application built in the RADICAL-EnTK API [10] to fully take advantage of the capabilities offered by the RCT suite.

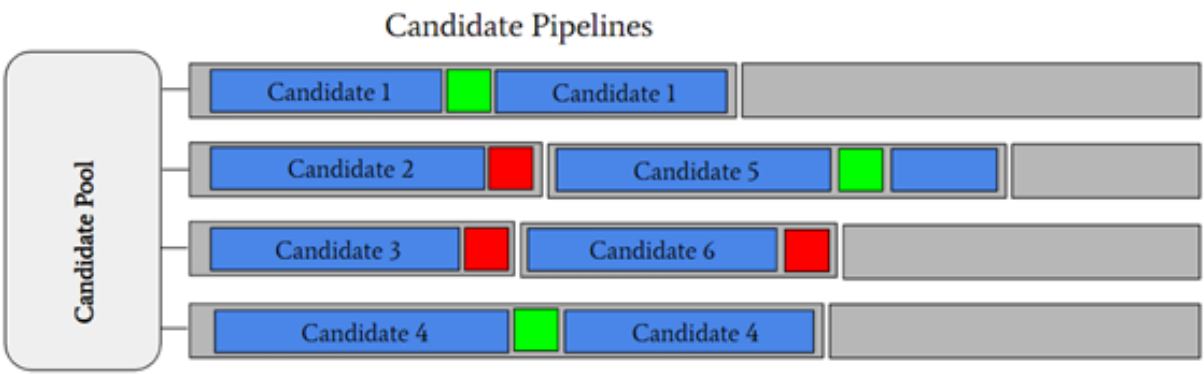

**Figure 1:** PACE[2] Schematic. Candidates set up a priori are drawn sequentially from the Candidate Pool and simulated in concurrent Pipelines. Each Pipeline performs an MD phase (blue) followed by an analysis phase. If the analysis task returns a "compliant" signal (green), the Pipeline extends into an additional MD phase. A "non compliant" (red) signal eliminates the Candidate by terminating the Pipeline and spawns a new Pipeline with the next Candidate.

PACE[2] is a workflows-based framework for simulating and investigating biomolecular material systems. A predefined number of different systems, such as different molecular chemistries or compositions at different physical conditions (temperatures, pressure, concentrations, or pH) are prepared and set up as Candidates in a Candidate Pool. A Candidate Pool is a set of prepared data, e.g., input structures, topologies, or runtime parameters. In practice, it is a sandbox that encompasses all the necessary input files. As many Pipelines as desired are instantiated, say, k in number, and Molecular Dynamics simulations of the first k Candidate systems are simultaneously spawned. Based on available resources, typically a single Pipeline is assigned to a node of a HPC platform. These Pipelines may be handled in one of two modalities: Elimination, or Extension. In the Elimination mode, the biomolecular systems are screened for the desired characteristics, and either allowed to propagate to the high resolution MD phase or terminated based on the screening analysis signal. In the Extension mode, different protocols are applied sequentially to the biomolecular system depending on the end state of each phase. This entire set of Pipelines may then be replicated for statistical significance. Figure 1 shows a schematic representation of the PACE$^2$ workflow. The blue blocks represent the Molecular Dynamics simulations associated with the Candidates, whereas the green and red blocks show the Analysis Phase. The green blocks represent an Analysis Phase that have returned a positive (Boolean "1") signal, and the red blocks, a negative (Boolean "0") signal. The Analysis can in principle be anything: from an explicit determination of simulation parameters to using statistical or analytical models to make decisions.

*PACE$^2$ Implementation*

The object-oriented design for PACE$^2$ is facilitated through two classes: Candidate and CandidateManager. These two classes are used by the PACE$^2$ top-level driver script to invoke all PACE$^2$ functionalities. Using these two classes, PACE$^2$ is able to carry out automated compliance-based elimination and extension through implementation of the radical.entk API.

At an abstract level, an instance of a Candidate class represents a simulation candidate, containing a set of Phases: pre-MD (preMD) phase, an MD phase, and an Analysis (AN) phase. Each iteration of these three phases is termed as a "Cycle". The preMD phase specifies any preprocessing required prior to running the MD phase. The MD phase performs the Molecular Dynamics simulations to compute the trajectories associated with the system dynamics. The AN phase computes a compliance metric for the simulations to facilitate the modification of the Candidate task graphs. The Candidate class does not inherently run each of these phases, but rather specifies their behaviors and interactions. Each of these phases is composed of Stages (which is a Python Class inherited from *radical.entk*).

Each simulation candidate, i.e. each Candidate object, is an ordered list of these phases and their corresponding Stages for the given unique candidate. To facilitate the elimination and extension of Candidates, the Candidates invoke the function *_extend_pipeline()*, which is a class method of the CandidateManager class to check for the compliance of each simulation Candidate, extending additional Pipeline Cycles or terminating the Pipeline as necessary. The Candidate specifications, i.e., all runtime details required to run a Candidate to completion are provided in a Python dictionary, and explicitly specify the following: (i) the simulation basename, (ii) the total number of Candidates, (iii) the number of Pipeline cores, (iv) the executable for the Molecular Dynamics simulation engine, (v) an executable for any necessary pre-processing, (vi) the arguments for the Molecular Dynamics simulation, (vii) the executable for the analysis, (viii) the arguments for the analysis and (ix) the maximum number of cycles as a threshold to prevent infinite cycling. After successfully defining the task graph of each Candidate through instances of the Candidate class, the top-level PACE$^2$ driver script creates a CandidateManager object with the details of the simulation ensemble specified to run all of the Candidates provided.

The object oriented design also allows for abstraction of kernel runtime details, i.e., the details of the kernel code that runs the Molecular Dynamics simulations or analysis. Since the task objects that own these kernels are agnostic to the kernels themselves, PACE$^2$ provides complete kernel agnosticity. This is facilitated by the configuration file that allows for the specification of all required parameters for running any user-chosen kernel in the form of a Python dictionary. This means that the MD phase can run any Molecular Dynamics engine of the user's choice. Furthermore, the framework does not necessitate that the kernel be a Molecular Dynamics engine: engines associated with other simulation techniques such as Monte Carlo may be employed to pursue specific scientific problems. Similarly, the "Analysis" kernel could be any standalone script or pre-compiled code that already resides on the target resource's storage disk.

An important step in determining the task graph is dependency resolution. Not all tasks that are known to reside on the task graph will have their dependencies met *a priori,* and these dependencies must be resolved. For example, a Molecular Dynamics task that is known to exist at a future Cycle may not yet have its input files generated. Once these files are generated in the prior Cycle, they must be appropriately moved to the correct sandboxes. PACE$^2$ makes this determination using the analysis kernel, and communicates the resolved dependencies to EnTK, which then creates symbolic links to the appropriate files at runtime.

The backbone of the implementation of PACE$^2$ is the Candidate Python class. The class consists of class methods to create the Candidate container (i.e., an EnTK Pipeline object), and methods to populate the Pipeline with the MD and Analysis phases (which are also EnTK objects: Stages and Tasks). The class also provides a function to perform the necessary decision making to extend or eliminate the Candidate. The Candidate Manager class creates as many containers as are necessary to host the Candidates of interest, and manages the spawning and execution of each Candidate. Finally, the top level driver script acts as the launcher for the PACE$^2$ application, allowing it to collect user input, spawn the Candidate Manager object, and terminate the application when all Candidates have completed execution. Table A.2 details the various components of the PACE$^2$ implementation. These components interact with each other as laid out in the table to enable the necessary execution containerization and dependency resolution.

## RESULTS AND DISCUSSION

The PACE$^2$ implementation is used to examine the extensive phase diagram of two biomolecular materials formed via dynamic interfacial phenomena, namely self-assembly and membrane-fusion. The first biomolecular material is formed via the self-assembly of two distinct peptide sequences. The molecular composition of these materials dictates their morphology. The second biomolecular material is formed via the fusion of a membrane encompassing a mixture of phospholipids and dendron-grafted amphiphiles. The molecular composition of these materials can yield stable or unstable vesicles. The application of PACE$^2$ to the two materials systems is discussed in two case studies.

For eac h use case, the PACE$^2$ framework has been tested on both local desktop computing (i.e., local hosts) and external HPC (external hosts) resources to demonstrate the portability of the framework. The tests on the local hosts for each Candidate do not run Molecular Dynamics simulations of the entire dynamical process but use the final structures from prior studies [3,4] to run short Molecular Dynamics simulations whose outputs are tested by suitable analysis routines. This was done for the sake of computational efficiency: the simulations were not being tested for their outcomes, but being used as workloads to test the Elimination/Extension capabilities of PACE$^2$. In contrast, the tests on the external hosts for each Candidate ran Molecular Dynamics simulations resolving entire dynamical processes whose outputs are tested by suitable analyses.

*Case 1: Bi-component Peptide Coassembly*

The morphological diversity of a biomolecular material with two peptide sequences is vast. The sequence of the two peptides dictates its structure, which in turn impacts the morphology of the equilibrium self-assembled nanostructure. We explore such a system as a use case for the PACE$^2$ framework: a bicomponent system of coassembling peptides. The peptides in question are diphenylalanine (FF) and phenylalanine-asparagine-phenylalanine (FNF), a system studied previously and reported to exhibit a vast morphological diversity [3] (see Figure 2).

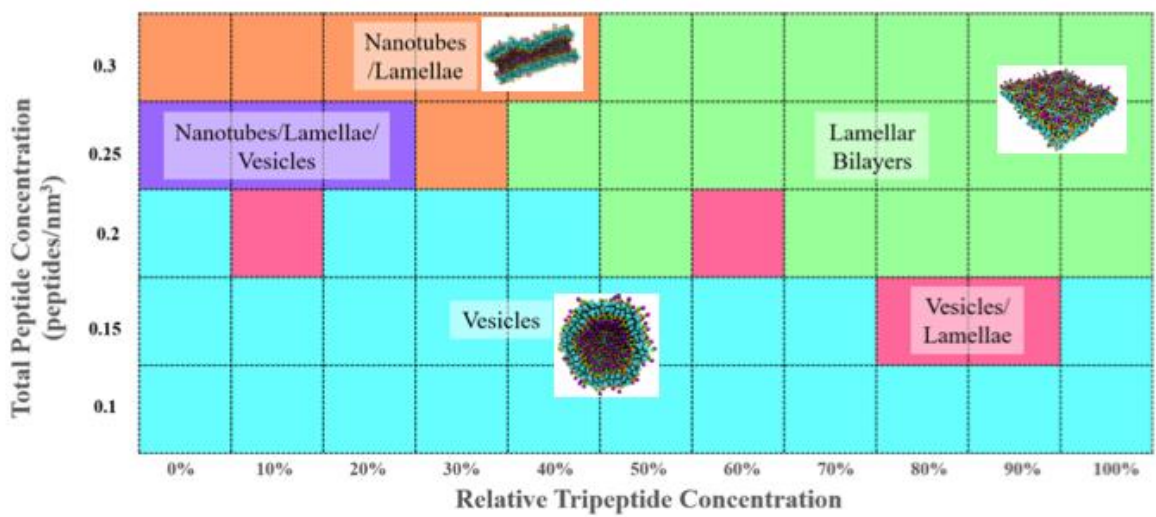

**Figure 2:** Phase map of bicomponent peptide coassembly of FF and FNF as a function of relative tripeptide and total peptide concentrations. [3]

A system such as the one described in this case study is particularly amenable to the PACE$^2$ framework. It involves a large ensemble of simulations, and requires the ability to discern simulations that yield desirable morphologies from those that do not, without requiring explicit user intervention. This is facilitated by a convolutional neural network (CNN) set up for image recognition, as discussed ahead.

A CNN [11] driven image recognition kernel is adopted to discriminate between the three possible nanostructures of FF-FNF peptide systems. This approach is an image classification problem. CNN driven image recognition has been shown to be particularly effective.[12] The technique involves "flattening" (i.e., reducing the dimensionality) of the input coordinate set of the supramolecular assembly from the trajectory file using a fixed perspective. The flattened coordinates are then used to generate an image file. This process converts the three dimensional coordinate data to a two dimensional image which may be operated upon by the CNN. An alternative approach would have been to employ a three dimensional neural network [13] which treats a set of three dimensional coordinates as several "temporally" separated two dimensional "slices" to scan all three dimensions. However, the former method was adopted due to performance concerns as well as the fact that the structural differences between the morphologies are sufficiently distinct so as to not require a full three dimensional scan.

The kernel was set up with a converter script to generate .png image files 480 by 640 pixels in size from the GROMACS output .gro files. This was done by plotting the *x, y* and *z* coordinates of the output coordinate file by fixing the perspective using Python's matplotlib library. This image was then passed on to the CNN. Figure 3 shows the images that were input into the CNN.

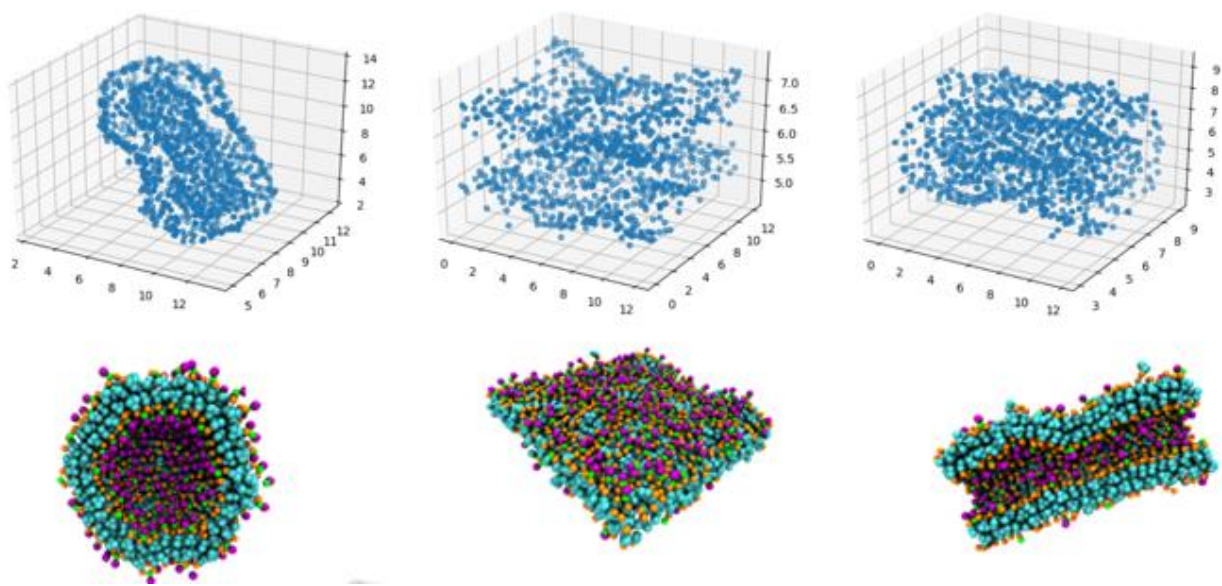

**Figure 3:** Images of a vesicle, lamellae and nanotube (Left to Right) generated by the Molecular Dynamics engine (bottom) and the pre-processing module of the neural network (top).

The images generated by the matplotlib driven preprocessing module were then passed to the neural network. For the training of the CNN, a labeled set of configurational data for the FF-FNF systems was used to generate the corresponding images. However, during runtime prediction of the supramolecular structures within PACE[2], the images were generated as the Candidates completed their Molecular Dynamics simulations. Figure 4 shows the scheme for the data flow from the Molecular Dynamics step to the machine learning step until the shape of the assembly is characterized.

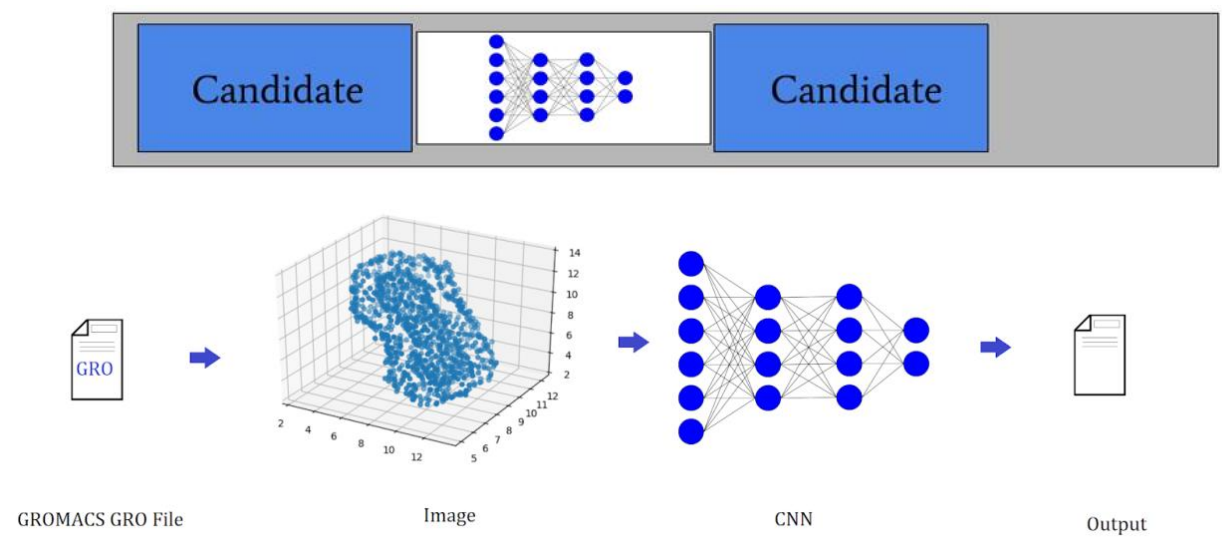

**Figure 4:** (Top) Setup of a Pipeline. The Candidate simulation runs for a predefined number of timesteps, and is operated upon by a CNN that reads the output of the simulation. (Bottom) Setup of the CNN kernel. A GROMACS .gro file is first used to generate an image that is then input into the CNN, which generates an output string and writes it to a file.

The Image recognition kernel was set up using TensorFlow. [14] The model was configured as a 4 layer CNN with one input layer, one output layer and two hidden layers. The input layer has 921600 neurons, three for each pixel of the image (three to correctly parse the RGB values). This is followed by Convolve() and MaxPool() operations. The two Convolution layers are set up with 9 neurons (3 x 3 setup). Max pooling is performed by a 2 by 2 kernel, i.e., the sub-matrices selected from the input matrix are 2 by 2. These operations are then repeated for further filtration and extraction of features. The output is a single string that reports if the image initially input into the network was a nanotube, vesicle or lamella.

The model was trained using 300 images each of vesicles, lamellae and nanotubes, and validated using 100 images of each morphology. Figure 5 shows the validation plot of the trained model. After 7 epochs, the accuracy of the prediction in the validation sets goes to ~1 in five independent training runs. Therefore, ten epochs is considered to be reasonable for good feature extraction. Extending the training further risks overfitting. [15]

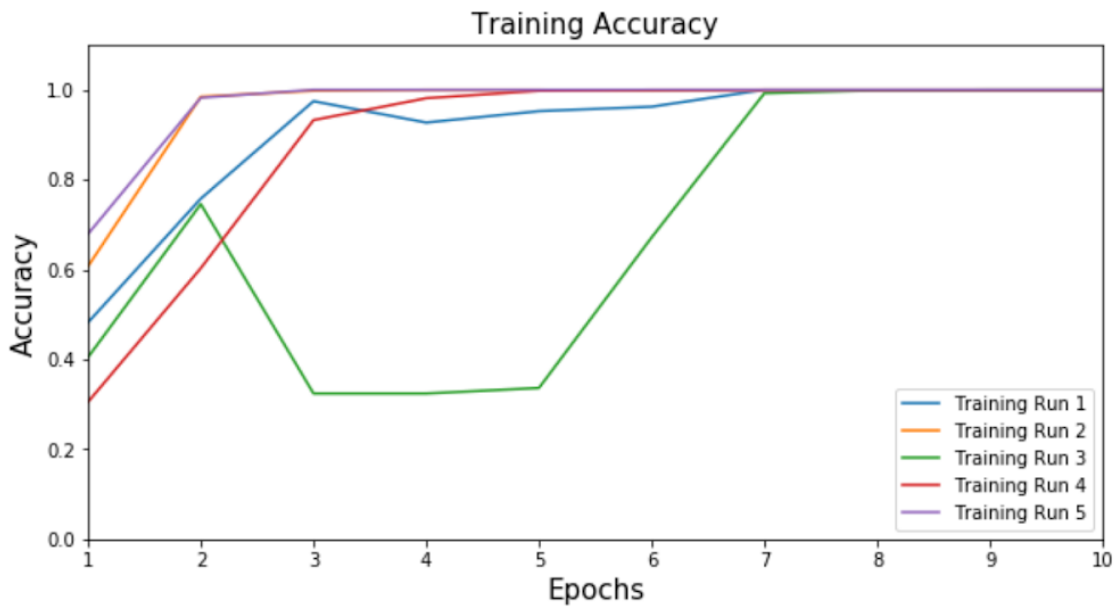

**Figure 5:** Training and validation of the CNN model. The training set consisted of 300 images each of the three nanostructures. The remaining 100 images were used to validate the model after each epoch. Five independent training runs were performed. All runs converged to a ~1.0 accuracy after 7 epochs.

The CNN model was trained as described in the *Methods* section. The trained model was then employed as the Analysis kernel in PACE$^2$ (see Figure 6). For testing PACE$^2$ framework on local hosts, a set of 9 Candidates was set up with molecular systems from a prior publication. [3] Each Candidate Pipeline was selected with a specific set of parameters reflecting the total concentration and relative tripeptide concentration as shown in Table 1. The "Detector", i.e. the analysis kernel, was set up to detect nanotubes, vesicles, or lamellae. The starting structures were chosen to be as close to the final stable self-assembled structures as possible. The simulations were run within PACE$^2$ for 5000 timesteps, at the end of which the outputs were put through the Analysis Kernel. This Analysis Kernel then triggered the Extension/Elimination signal to the Pipeline. If the Pipeline was extended, it was run for 1000 additional timesteps. Else, it was terminated.

| CandidateID | Molecular Comp | | Detector (CNN) | Timesteps before CNN Analysis | Extension | Comparison with OBC |
|---|---|---|---|---|---|---|
| | Total Conc | Relative Tripeptide | | | | |
| 1 | 0.3 | 10% | N | 5000 | Extended | Agreement |
| 2 | 0.1 | 10% | V | 5000 | Terminated | Agreement |
| 3 | 0.3 | 60% | L | 5000 | Terminated | Agreement |
| 4 | 0.3 | 20% | N | 5000 | Terminated | Agreement |
| 5 | 0.1 | 20% | V | 5000 | Extended | Agreement |
| 6 | 0.3 | 70% | L | 5000 | Extended | Agreement |
| 7 | 0.3 | 30% | N | 5000 | Terminated | Agreement |
| 8 | 0.1 | 30% | V | 5000 | Terminated | Agreement |
| 9 | 0.3 | 80% | L | 5000 | Terminated | Agreement |

**Table 1:** Simulation setup to validate Pipeline extension/elimination protocol for case 1 on local hosts. The starting structures in each Pipeline were chosen to be as close to the final structures as possible. The setup tested the extension behavior of the Pipelines as dictated by the "detector", i.e., the CNN-based Analysis kernel in PACE$^2$ trained to detect the presence of either Nanotubes, Vesicles, or Lamellar Bilayers.

The extension behavior of the Pipelines was found to have a 100% success rate on the self-assembled nanostructures. The CNN correctly identified the resulting nanostructure, successfully communicated the correct signal to PACE[2], and triggered the correct adaptation to the Pipeline (namely, extend/eliminate).

*Case 2: Membrane Fusion*

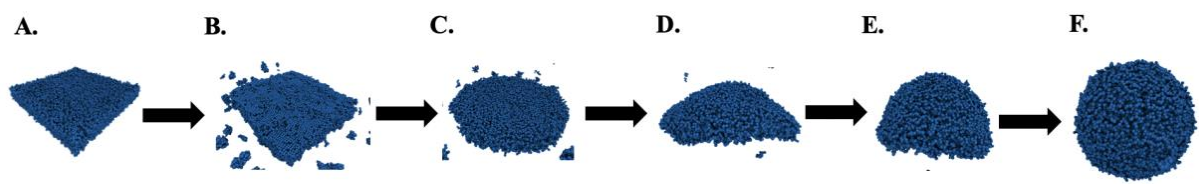

**Figure 6.** A-F shows the transition of a bilayer (A) to a vesicle (F). Blue beads are DPPC lipids. PDAs are not shown for clarity.

The formation of vesicles via different self-assembly pathways has been investigated by theory [16–19] and experiments. [20] In particular, the bicelle-to-vesicle transition has been extensively studied as one of the possible self-assembly mechanisms. [21] Here, a disk-like membrane called a bicelle (Figure 6.C) fuses its edges to form a vesicle (Figure 6.F). Hence, this phenomenon is also termed "membrane fusion". In a previous study, [4] a phase space of dendronized vesicles is reported (Figure 7). These vesicles consist of polyamidoamine dendron-grafted amphiphiles (PDAs) and dipalmitoyl-*sn-glycero*-3-phosphocholine (DPPC) lipids. The phase space of these vesicles is explored as a function of PDA generation and concentration. The green region in Figure 7 represents the design conditions that generate stable vesicles. On the other hand, the red region represents design conditions that do not generate stable vesicles. Since membrane fusion simulations consume significant computing resources, the PACE[2] framework can terminate simulations that do not generate stable vesicles.

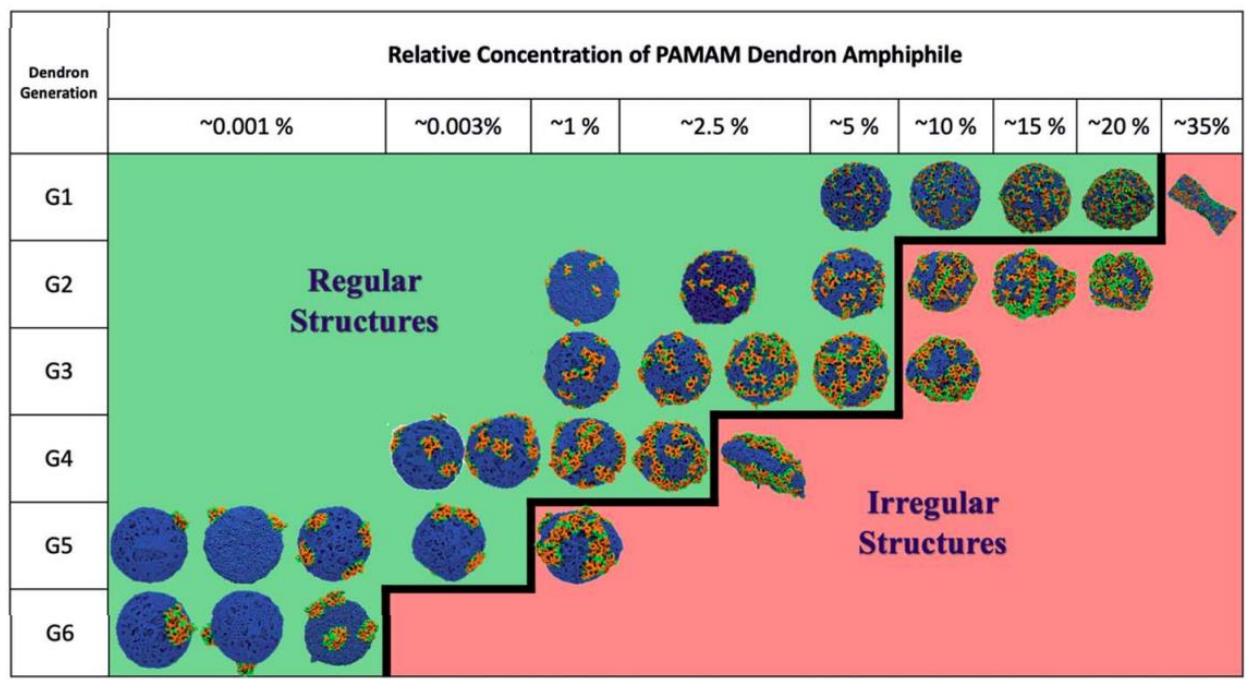

**Figure 7**. Phase map of dendronized vesicles generated via membrane fusion. It is plotted as a function of PDA generation and relative concentration. Reproduced from Ref. [4] with permission from the Royal Society of Chemistry.

To generate dendronized vesicles, the process of membrane fusion begins with an equilibrated bilayer consisting of PDAs and DPPC lipids. The equilibrated bilayer (Figure 6.A) is placed at the center of a large simulation box encompassing water, with the edges of the bilayer exposed to water. This configuration generates unfavorable interactions between the hydrophobic moieties in the bilayer and water. The bilayer minimizes these interactions by transitioning to a disk-like structure (namely, a bicelle) and partially fusing its edges (Figure 6.C). Next, the bicelle further minimizes the unfavorable interactions by transitioning to a bowl-like structure (see Figure 6.E), enroute to forming a stable vesicle (Figure 6.F).

At high concentration of PDAs, the membrane fusion process does not yield stable vesicles,

i.e., the membranes do not fuse their free edges to form a vesicle. The PACE$^2$ framework is used to eliminate these simulations from an ensemble of simulations. In this way, computational resources are optimally employed to simulate only membranes that are able to successfully form stable vesicles. Like case 1, a method to discern simulations that yield desirable morphologies is required. This is facilitated by a K-means clustering code that can determine if the membrane has fused or not.

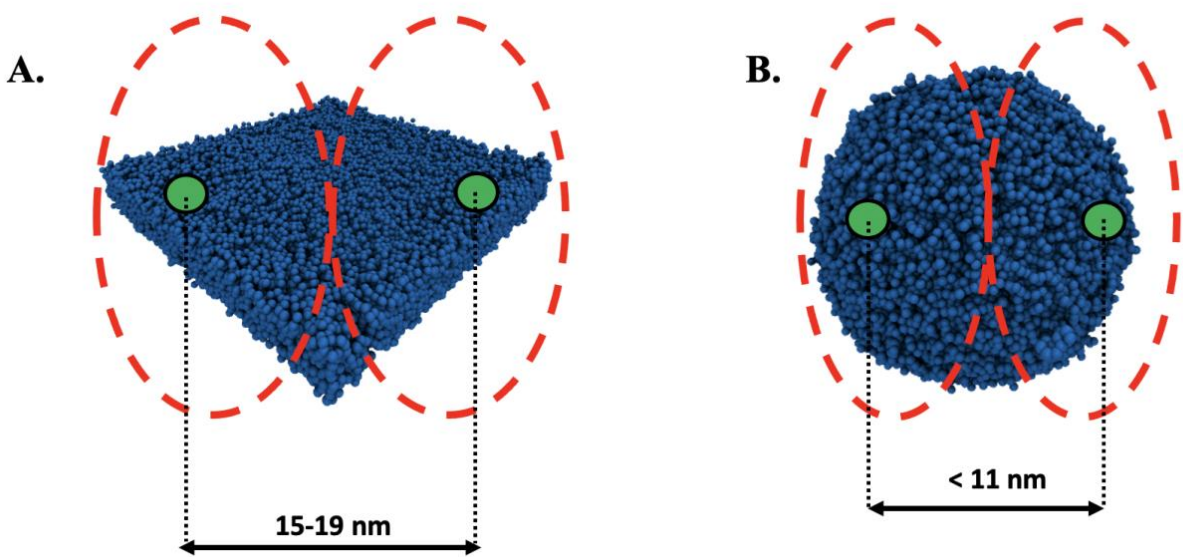

**Figure 8.** Images of a flat membrane (A) and a spherical vesicle (B). The red dashed lines divide the assemblies in (A) and (B) into 2 regions of equal size. The green circles represent the center of mass of the clusters.

The k-means clustering algorithm [22,23] is used to divide the membrane into two clusters of equal size (red dashed lines in Figure 8). The k-means code provides the cartesian coordinates for the center of mass of each cluster (green circles in Figure 8). The Euclidean distance between the center of masses of the two clusters is considered to be the characteristic length of the structure. The characteristic length of flat membranes in this class of systems is between 15 and 19 nm.[4] On the other hand, the characteristic length of a spherical vesicle is less than 11 nm.[4] Hence, the characteristic length is used to distinguish between desirable (membranes fusing to yield stable vesicles) and undesirable (membranes that do not fuse their edges to form stable vesicles) simulations.

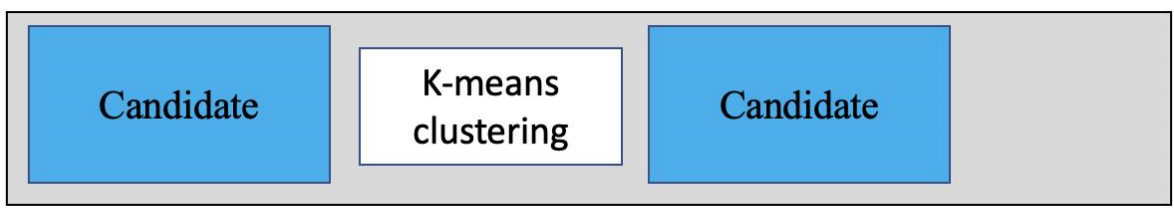

**Figure 9:** Setup of the K-means clustering kernel in a Pipeline. The GROMACS output coordinate file from the first Stage is provided as an input to the K-means clustering code, which generates an output string and writes it to a file. Based on the output, the simulation is either extended or eliminated.

The k-means clustering code was employed as the Analysis kernel in PACE$^2$ framework (see Figure 9). To validate the adaptive functionality on local hosts, membrane fusion simulations for 10 Candidates (with a specific generation and concentration of PDAs) were conducted using the PACE$^2$ framework (see Table 2). The analysis kernel was set up to determine the characteristic length of the structure. Like case 1, the starting structures were run within PACE$^2$ framework for 5000 timesteps, and the output structures were passed to the Analysis kernel. If the characteristic length of the structure was less than or equal to 11 nm, an extension signal was triggered, resulting in another simulation spanning 5000 timesteps. If the characteristic length of the structure was more than 11 nm, the Pipeline was terminated.

| CandidateID | PDA parameters | | Detector (K-means) | Timesteps before K-means Analysis | Extension | Comparison with Ref Nanoscale |
|---|---|---|---|---|---|---|
| | Generation | Relative Concentration | | | | |
| 1 | 1 | 10% | 9.5 nm | 5000 | Extended | Agreement |
| 2 | 1 | 25% | 15.5 nm | 5000 | Terminated | Agreement |

| 3 | 2 | 4% | 10.4 nm | 5000 | Extended | Agreement |
| 4 | 2 | 7% | 16.6 nm | 5000 | Terminated | Agreement |
| 5 | 3 | 3% | 10.4 nm | 5000 | Extended | Agreement |
| 6 | 3 | 6% | 17.6 nm | 5000 | Terminated | Agreement |
| 7 | 4 | 1.4% | 9.8 nm | 5000 | Extended | Agreement |
| 8 | 4 | 3% | 17.3 nm | 5000 | Terminated | Agreement |
| 9 | 5 | 0.004% | 10.2 nm | 5000 | Extended | Agreement |
| 10 | 5 | 0.01% | 18.4 nm | 5000 | Terminated | Agreement |

**Table 2**: Simulation setup to validate Pipeline extension/elimination protocol for case 2. The setup tested the extension behavior of the Pipelines as dictated by the "detector", i.e., the K-means kernel that determines if the membrane has fused its free edges to form a vesicle or not.

Like case 1, the extension capability of the PACE$^2$ framework worked correctly for all of the 10 Candidates. The analysis kernel (namely, k-means clustering) measured the characteristic length of the structure and communicated the correct signal to PACE$^2$, i.e., to either extend or terminate the Pipeline.

## CONCLUSIONS

Computational investigations of the dynamical processes underlying the formation of biomolecular materials under a variety of compositions and external conditions can be efficiently performed using ensemble MD simulations. This approach faces two challenges: the large number of independent ensembles required to investigate the composition phase space, and the difficulty in predicting the formation of the material and the duration of each simulation. These challenges can be addressed by management strategies which include intermittent inspection, variable sampling, premature termination and extension of individual Molecular Dynamics simulations. Furthermore, the management strategies can be automated to facilitate the long-term management of a large ensemble of Molecular Dynamics simulations.

To this end, a high-throughput workflows-based computational framework for biomolecular material simulations is developed and its applications are discussed. The computational framework, PACE$^2$, consists of a set of Simulation-Analysis Pipelines, each comprising an arbitrary number of Cycles, i.e., Simulation-Analysis Loops. The Pipelines can be replicated for concurrent processing of multiple Candidates simultaneously. The capabilities of the PACE$^2$ framework are demonstrated through two use cases: peptide coassembly and membrane fusion. The peptide coassembly case demonstrates the operation of the framework in its "Elimination" protocol, i.e., stopping undesirable simulations and extending desirable ones for a limited number of Cycles. In contrast, the membrane fusion case demonstrates the "Extension" protocol, i.e., extending a simulation for multiple cycles until a desired end state is achieved. These two cases are not mutually exclusive: it is possible to design a workload that combines these two protocols in any arbitrary manner, such as eliminating unfavorable Pipelines while extending favorable ones until the desired endstate is achieved. The operability of the framework has been demonstrated on a local machine as well as a HPC platform. The source code and examples can be found at [24].

In the future, the PACE$^2$ framework will be extended to address more generalized workflow systems. In addition, other computational methods and analysis tools prevalent to the identification of composition-structure-property relations in diverse classes of materials, such as polymeric materials, composites, alloys, ceramics and glasses, can be potentially integrated into the framework.

## ACKNOWLEDGEMENTS

Portions of this work were performed using high performance computing resources provided by the National Science Foundation XSEDE allocation DMR-140125. The authors gratefully acknowledges financial support from the National Science Foundation CAREER award DMR-1654325, awards OAC-1547580, NSF OAC-1835449, and DMREF DMR-2118860. The authors also thank Dr. Mikhail Titov for help debugging RADICAL-Cybertools on XSEDE resources.

# APPENDIX A: TERMINOLOGY

| Term | Definition |
|---|---|
| Candidate | An abstraction consisting of all inputs and outputs of a specific simulation. From the user perspective, each Candidate is initially represented by a directory with the required input files. Inputs include model definitions or force fields, initial structures, and runtime parameters. A Candidate is executed in a Pipeline as a sequence of Stages. See Figure 1. |
| Candidate Pool | A set of Candidates prepared by the user. Often Candidates represent related simulations which take on different values of an instrumental variable, although this is not a requirement. See Figure 1. |
| Cycle | A compound process block usually comprising an MD Phase and an Analysis Phase. A complete Cycle produces a text output containing a 0 or 1, which PACE$^2$ uses to determine whether the parent Pipeline is extended or terminated. See Figure 1. |
| Phase | The functional process block of a typical PACE$^2$ workflow in which a simulation is run and then analyzed. For example a usual MD Phase consists of: 1) environment initialization (e.g. module load x) 2) Molecular Dynamics engine initialization (i.e. gmx grompp) 3) Molecular Dynamics execution (gmx mdrun). The Analysis Phase usually includes the execution of an arbitrary analysis script with an output of 0 or 1. See Figure 1. |
| Pipeline [EnTK] | A sequence of Stages, which can be dynamically ordered and bound to a reserved compute resource. Pipelines are managed by PACE$^2$ and are not explicitly invoked by the user. Note this is an EnTK entity. |
| Sandbox [EnTK] | A directory created by EnTK which contains all of the input files required for task execution as well as all task outputs. Sandboxes are managed by EnTK and not explicitly invoked by PACE$^2$ or the user. Note this is an EnTK entity. [Ref entk docs – sandbox only seems to appear as task.sandbox] |
| Stage [EnTK] | A set of independent tasks to be executed concurrently. Stages are managed by PACE$^2$ and are not explicitly invoked by the user. Note this is an EnTK entity. |

| Task | Unitary components of a computational toolchain, typically but not always equivalent to an EnTK Task. In the context of PACE$^2$, a Task is effectively a single call to a command line executable. See Figure 1. |
|---|---|
| Task [EnTK] | An abstraction of a computational task that contains information regarding an executable, its software environment and its data dependences. Note this is an EnTK entity. |

**Table A.1:** Definitions of terms used in the manuscript, including reference definitions for relevant EnTK terms. [7,25]

| **candidate_manager.py** | | | | | | |
|---|---|---|---|---|---|---|
| **Function** | **Description** | **Inputs** | **Side effects** | **Return** | **Output (Abstraction level)** | **Comments** |
| *__init__* | Initialize Candidate Manager communications and interface with EnTK | hostname, port, username, password, resource dictionary, list of Pipelines | NA | NA | NA | |
| *run* | Instantiate with EnTK AppManager | Initialized Candidate Manager | NA | NA | NA | |
| **candidate.py** | | | | | | |
| **Function** | **Description** | **Inputs (Function Level)** | **Inputs(Abstraction Level)** | **Return** | **Output (Abstraction level)** | **Comments** |
| *__init__* | Initialize candidate | candidate specifications dictionary, candidate ID | NA | NA | NA | |
| *_create_candidate_pipeline* | Triggers the Pipeline management functions of the Candidate class | self (uses inputs specified at the Candidate object level) | All input files for MD simulations | Pipeline (EnTK object) | Pipeline (EnTK object) | |

| Function | Description | Inputs (Function Level) | Inputs (Abstraction Level) | Return | Output (Abstraction level) | Comments |
|---|---|---|---|---|---|---|
| _create_pre_md_stage | Creates pre_md Stage to run any pre-md preprocessing required by the MD kernel (GROMACS specific) | self (uses inputs specified at the Candidate object level) | All input files for MD simulations | NA | NA | Contains in-function string generation routines to handle all input file pathing needed for the Grompp (GROMACS PreProcessor) |
| _create_md_stage | Creates MD stage and instantiates MD task. | self (uses inputs specified at the Candidate object level) | All input files for MD simulations | NA | NA | Contains in-function string generation routines to handle all input file pathing needed for the mdrun (GROMACS MD kernel) |
| _create_analysis_stage | Creates analysis stage and instantiates analysis task | self (uses inputs specified at the Candidate object level) | Output of MD simulations | Stage (EnTK Object) | Stage (EnTK Object) | Contains in-function string generation routines to handle all input file pathing needed for the analysis script. Analysis Stage object needs an explicit return because it is referenced in the function that creates it. |
| _extend_pipeline | Makes pipeline extension/elimination decision | self (uses outputs generated by the analysis task and internally linked into the task sandbox by RADICAL-Pilot) | Output of Analysis Stage | NA | NA | Uses the output of the analysis Stage to make a decision to either continue (*Extend*) or kill (*Eliminate*) the pipeline that triggers this |
| pace.py | | | | | | |
| **Function** | **Description** | **Inputs (Function Level)** | **Inputs(Abstraction Level)** | **Return** | **Output (Abstraction level)** | **Comments** |
| *main* | Initialize the required number of Pipelines | NA | NA | NA | NA | |
| *read_jsons* | Creates a simulation and resource dictionary | Simulation and resource json files | NA | Simulation and Resource Dictionary | NA | Reads all inputs arguments required to run the ensemble of simulations. |

**Table A.2:** Functions implemented in PACE[2] to enable remote execution of task graphs.